\documentclass[twocolumn,5p]{elsarticle}

\usepackage{amsfonts, amssymb, amsmath}

\usepackage{graphics}
\usepackage{graphicx}
\usepackage{epsfig}
\usepackage{amssymb}
\journal{Physica B}
\usepackage{wasysym}

\usepackage{color}
\definecolor{darkblue}{rgb}{0,0,0.6}
\definecolor{darkgreen}{rgb}{0,0,0.6}
\definecolor{darkred}{rgb}{0.6,0,0}
\usepackage[colorlinks=true,urlcolor=darkblue,citecolor=darkblue,linkcolor=darkred,hyperfootnotes=false]{hyperref}

\newcommand{\argc}[1]{\left[#1\right]}
\newcommand{\arga}[1]{\left\lbrace #1\right\rbrace }
\newcommand{\argp}[1]{\left( #1 \right)}
\newcommand{\valabs}[1]{\vert #1\vert}
\newcommand{\moy}[1]{\left\langle  #1 \right\rangle }

\begin{document}

\begin{frontmatter}




\title{Disordered Elastic Systems and One-Dimensional Interfaces}

\author[authoradressGeneve]{Elisabeth Agoritsas\corref{EA} }
\cortext[EA]{Corresponding author (e-mail: \texttt{Elisabeth.Agoritsas@unige.ch})}
\author[authoradressParis]{Vivien Lecomte}
\author[authoradressGeneve]{Thierry Giamarchi}

\address[authoradressGeneve]{DPMC-MaNEP, University of Geneva, 24 Quai Ernest-Ansermet, 1211 Geneva 4, Switzerland}
\address[authoradressParis]{Laboratoire Probabilit\'es et Mod\`eles Al\'eatoires, UMR CNRS 7599, Universit\'es Paris VI et Paris VII, Site Chevaleret, \\ 175~rue du Chevaleret, 75013 Paris, France}

\begin{abstract}

We briefly introduce the generic framework of Disordered Elastic Systems (DES), giving a short `recipe' of a DES modeling and presenting the quantities of interest in order to probe the static and dynamical disorder-induced properties of such systems.
We then focus on a particular low-dimensional DES, namely the one-dimensional interface in short-ranged elasticity and short-ranged quenched disorder. Illustrating different elements given in the introductory sections, we discuss specifically the consequences of the interplay between a finite temperature $T>0$ and a finite interface width $\xi>0$ on the static geometrical fluctuations at different lengthscales, and the implications on the quasistatic dynamics.
\end{abstract}

\end{frontmatter}


\section{Introduction}
\label{section-introduction}

Could some features of experimental systems as dissimilar at a microscopic level
as superconductors, magnets, ferroelectrics, fluids, paper, or two-dimensional electron gases,
be described by the same equations at a macroscopic level?
All those systems may actually display emergent structures
such as \textit{interfaces} (e.g. ferroelectric \cite{tybell_2002_PhysRevLett89_097601,paruch_2005_PhysRevLett94_197601,pertsev_2011_JApplPhys110_052001} or ferromagnetic \cite{lemerle_1998_PhysRevLett80_849,repain_2004_EurPhysLett68_460,metaxas_2007_PhysRevLett99_217208} domain walls,
contact line in wetting experiments \cite{moulinet_2002_EurPhysJE8_437}
or propagating cracks in paper and thin materials  \cite{santucci_2007_PhysRevE75_016104})
or \textit{periodic systems} (typically vortex lattices in type-II superconductors \cite{blatter_1994_RevModPhys66_1125},
classical \cite{coupier_2005_PhysRevE71_046105} or quantum \cite{giamarchi_2004_arXiv:cond-mat/0403531} Wigner crystals,
or electronic crystals displaying charge or spin density waves \cite{gruner_1988_RevModPhys60_1129,brazovskii_2004_AdvPhys53_177}).

One can either describe them using \textit{ab initio} predictions combined to a Landau approach, where two phases compete with each other at their common boundary
(the complexity of a numerical approach increasing considerably with the system size),
or rather take a radically opposite point of view by skipping the specific microphysics and focusing exclusively on the boundary, defined by the shift of the order parameter.
Such an emergent structure can then be described as a fluctuating manifold or periodic system supported by a disordered underlying  medium, in the generic framework of \textit{disordered elastic systems} (DES).

Thereafter we recall briefly the basic features of DES by giving first a short recipe of a DES model based on two competing physical ingredients: elasticity and disorder, blurred by thermal and/or quantum fluctuations.
Then we list the main observables of interest in order to probe the disorder-induced metastability present in those systems, and to address the two main questions which arise regarding their resulting glassy properties:
what can we learn by the study of its statics \textit{versus} its dynamics, first via the characterization of its geometrical fluctuations and secondly via its response to an external force?

Indeed, from the 1970s' and Larkin's work \cite{Larkin_model_1970-SovPhysJETP31_784}, we know that there could not exist a perfectly ordered solid in presence of disorder, so how does the addition of disorder change the nature of a pure system?
Finally we focus on a particular low-dimensional DES and study the static geometrical fluctuations of a one-dimensional interface, via an analysis of the interplay between thermal fluctuations and a finite width of the interface in its roughness.

Those short notes are not meant to be exhaustive, but rather to give a pedagogical and somehow practical introduction to the field, aimed at theoreticians but also at experimentalists who might be interested in DES modeling.
We focus essentially on the case of \textit{interfaces}, but most concepts can be extended to \textit{periodic systems}, and more details and references can be found for example starting from the existing reviews \cite{brazovskii_2004_AdvPhys53_177,giamarchi_2006_arXiv:0503437,TG_DES_Springer}.

\section{DES modelling: a recipe}
\label{section-DESrecipe}

In the generic framework of DES, very few physical ingredients are required in a minimal version of such a model. Thereafter we briefly sketch their concrete implementation for interfaces, but those considerations remain valid for periodic systems \cite{giamarchi_ledoussal_1995_PhysRevB52_1242}.

\subsection{Dimensionality and class of DES}

First of all one has to identify the \textit{dimensionality} of the system ($d$ being the internal dimension of the system, $m$ the number of its transverse components, and $D$ the dimension of its embedding physical space)
and whether it is a \textit{manifold} ($d+m=D$ with ${d=D-1}$ for interfaces) or a \textit{periodic system}.

For example, a 1D interface and a single vortex are manifolds respectively with ${(d=m=1,D=2)}$ and ${(d=1,m=2,D=3)}$, whereas an Abrikosov vortex lattice and a 3D Wigner crystal are periodic systems with ${(d=D=3,m=2)}$ and ${(d=m=D=3)}$.
Note that the dimensionality and the class of DES might actually change with respect to some parameters of the system, as it can been addressed experimentally e.g. for ferromagnetic domain walls (1D to 2D interfaces crossover) \cite{kab-jin_2009_nature09-458-740} or for vortices in superconductors (vortex lattice to individual vortices).

\subsection{Physical space of coordinates ${(\vec{z},\vec{x})}$}

The description of the physical space embedding the DES can then be split in two sets of coordinates:
${\vec{z} \in \mathcal{D}_z}$ denotes the \textit{internal} coordinates of the system (e.g. the position along a polymer or a given point in a lattice),
and ${\vec{x} \in \mathcal{D}_x}$ its \textit{transverse} coordinates.

For an analytical treatment of interfaces we typically assume that they live in an \textit{infinite} and \textit{continuous} physical space so ${\mathcal{D}_z \times \mathcal{D}_x}$ is taken as ${\mathbb{R}^d \times \mathbb{R}^m}$.
However, a physical or numerical realization of such a system
is always supported by a \textit{microscopically discrete} sublattice of parameter ${1/\Lambda}$, ultimately the crystal in a solid,
and moreover lives in a \textit{finite} box of typical size $L$ with boundary conditions which have to be defined (possibly periodic or free).
So in the comparison between analytical predictions and experimental or numerical results, corrections due to finite size effects and to the translation from the discrete to the continuous limits are \textit{a priori} expected.

Once the disorder is averaged out, a translational space-invariance is recovered, suggesting a description in Fourier space $q$ with an ultra-violet ${1/\Lambda}$ and an infra-red $1/L$ cutoffs,
which are always present in physical DES. However, from an analytical point of view, they are either irrelevant and thus skipped, or they are conveniently reintroduced in order to cure non-physical divergences in computations.

\subsection{Univalued displacement field $\vec{u}_{\vec{z}}$}

\begin{figure}
\includegraphics[width=8cm]{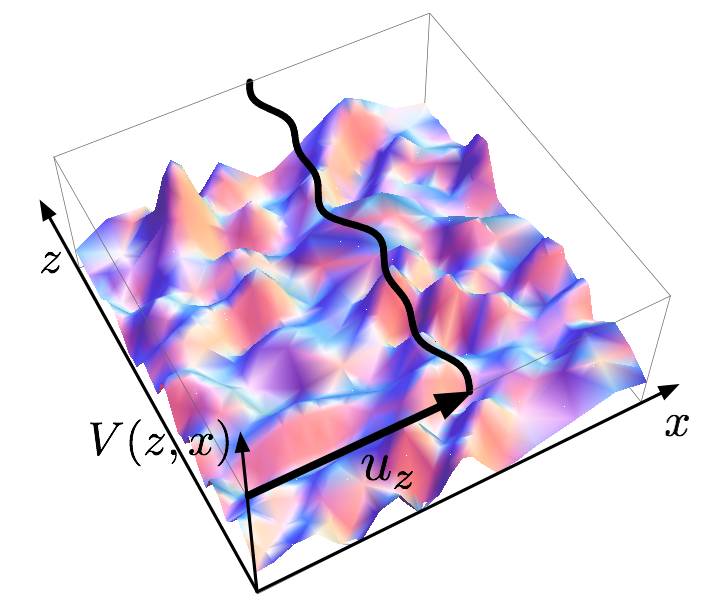}
\caption{Definition of the displacement field ${\vec{u}_{\vec{z}}}$ for a 1D interface (${d=m=1}$) superimposed over its surrounding smooth random potential ${V(\vec{z},\vec{x})}$ (in a weak disorder limit). ${\vec{u}_{\vec{z}}}$ is univalued only in the absence of bubbles or overhangs for interfaces.}
\label{fig-def-uz-dm1}
\end{figure}

In the absence of disorder, an elastic system tends to minimize its distortions,
thus would typically be flat for an interface or characterized by a single reciprocal vector for periodic systems.
A given configuration of a DES is characterized by a \textit{univalued displacement field} ${\vec{u}_{\vec{z}} \in \mathcal{D}_x}$ with respect to such an equilibrium configuration (e.g. flat or periodic) of the pure system as illustrated in Fig.\ref{fig-def-uz-dm1}.

The definition of this reference configuration is actually crucial but potentially tricky in experiments,
in particular in certain cases where the equilibrium configuration is not a straight line.
For example, how to define unambiguously the center and the mean radius of a `dotted' ferromagnetic domain if it is far from being perfectly circular?
Defects such as overhangs and bubbles for interfaces, or topological defects in periodic systems are still missing in the above DES description, since they hinder the definition of a univalued displacement field ${\vec{u}_{\vec{z}}}$, at the core of the concrete implementation of statistical averages both thermal and over disorder.

\subsection{Elasticity}

The elastic energetic cost of distortions ($\nabla_{\vec{z}} \vec{u}_{\vec{z}}$) is given by the elastic Hamiltonian
${\mathcal{H}_{\text{el}} \argc{u}}$.
For an interface, if the elasticity is \textit{short-ranged}, it is essentially proportional to the length of the interface ${\mathcal{H}_{\text{el}} \argc{u} \propto \int d^dz \, \sqrt{1 + (\nabla_{\vec{z}} \vec{u}_{\vec{z}})^2}}$
and in the \textit{elastic limit} of small distortions ${\Vert \nabla_{\vec{z}} \vec{u}_{\vec{z}} \Vert \ll 1}$ it reduces to the quadratic form (thus analytically user-friendly)
\begin{equation}
 \mathcal{H}_{\text{el}} \argc{u}
 = \frac{c}{2} \int_{\mathcal{D}_z} d^d z \cdot (\nabla_{\vec{z}} \vec{u}_{\vec{z}})^2
 \label{equa-Hel}
\end{equation}
with the elastic constant $c$ being the elastic energy per unit of length,
and that corresponds in reciprocal space to an elastic energy ${cq^2}$ per Fourier mode ${\vec{u}_{\vec{q}}}$.

If the elastic limit is broken, additional terms are \textit{a priori} needed in the perturbation expansion $\sqrt{1 + (\nabla_{\vec{z}} \vec{u}_{\vec{z}})^2}$.
Nevertheless the elasticity can also be effectively \textit{long-ranged}, as it is the case e.g. for contact lines in wetting experiments because of the fluid surface tension \cite{joanny_1984_JChemPhys81_552,ledoussal-wiese_2009_arXiv:0908.4001}.
The effective energy per Fourier mode is then $c\valabs{q}^{\mu}$ with $\mu \neq 2$, or the elastic Hamiltonian can include phenomenological additional terms as e.g. for the description of the out-of-equilibrium depinning regime (cf. section \ref{section-dynamics}) \cite{rosso-krauth_2001_PhysRevLett87_187002,kolton_2009_PhysRevB79_184207}.

\subsection{Disorder}

The `disorder' accounts for the effects of the inhomogeneities inherent to any real physical medium,
and it is defined as a stochastic variable ${V(\vec{z},\vec{x})}$ with a given statistical distribution ${\mathcal{P}\argc{V}}$ and the \textit{ad hoc} disorder average (denoted ${\overline{\mathcal{O}}}$ for an observable $\mathcal{O}$).
It adds a random part to the DES Hamiltonian ${\mathcal{H}_{\text{DES}}}=\mathcal{H}_{\text{el}}+\mathcal{H}_{\text{dis}}$ with the disorder Hamiltonian
\begin{equation}
 \mathcal{H}_{\text{dis}} \argc{u,V}
 = \int_{\mathcal{D}_z} d^d z \cdot V(\vec{z},\vec{u}_z)
 \label{equa-Hdis}
\end{equation}

First, if this stochastic variable has a dynamics much slower than the dynamics of the DES, the disorder is \textit{quenched} (as e.g. atomic terraces on thin epitaxial films) otherwise it is \textit{annealed} (as e.g. itinerant oxygen vacancies in superconductors).
That distinction essentially imposes the sequence of disorder and thermal averaging of observables in computations.

Secondly, disorder can either be dominated by a few individual pinning centers in the \textit{strong disorder} limit,
or by many weak impurities in the \textit{weak disorder} limit.
In the latter case, the collective behavior of the impurities conspire (by the central limit theorem) to give rise to a smooth random potential ${V(\vec{z},\vec{x})}$, as illustrated in Fig.\ref{fig-def-uz-dm1}, with a Gaussian distribution ${\mathcal{P}\argc{V}}$.
Note however that this `weak disorder' limit can be realized in experimental DES only because the disorder can actually vary on characteristic lengthscales which are much smaller than the lengthscales of interest of the DES description such as the lattice spacing of periodic systems (as e.g. for the 3D macroscopic Wigner crystal in colloids \cite{irvine-vitelli-chaikin_2010_nature_09620}).

${\mathcal{P}\argc{V}}$ is thus fully characterized by its two first cumulants,
namely its mean value ${\overline{V(\vec{z},\vec{x})} \equiv 0}$
and its variance
\begin{equation}
\overline{V(\vec{z},\vec{x})V(\vec{z}',\vec{x}')} = R_{\xi_z}(\vec{z}-\vec{z}') \cdot R_{\xi_x}(\vec{x}-\vec{x}')
 \label{equa-gendiscorr}
\end{equation}
The disorder is usually assumed to be uncorrelated along the internal direction $\vec{z}$ of the interface with ${R_{\xi_z}(\vec{z})=\delta(\vec{z})}$,
whereas the \textit{disorder correlator} ${R_{\xi_x}(\vec{x}})$ is the quantity extensively studied in the Functional Renormalization Group (FRG) approach of DES (cf. section \ref{section-methods}).
If this correlator decreases sufficiently fast, then the disorder is short-ranged or \textit{random-bond} (RB: the interface couples only locally to the surrounding disorder), else it is long-ranged or \textit{random-field} (RF: the system is sensitive to the disorder in all the physical space).
In the section \ref{section-1Dinterface} we examine specifically what are the consequences of a finite RB disorder correlation length ${\xi_x >0}$ on the static properties of a 1D interface.
Of course in an experimental system all types of disorder could be present, but the dominant disorder defines its universality class.

\subsection{Internal structure of DES}

At last, the internal structure of physical DES can be crucially relevant for its properties, and it is actually already partially encoded in the disorder correlator and its correlation lengths $\xi_z$ and $\xi_x$.

The implications of an \textit{internal degree of freedom}, as e.g. the phase shift in N\'eel \textit{versus} Bloch domain walls, are challenging theoretically, but such an additional physical ingredient in DES modelling might be needed to account for the physics of experimental systems, as it has been recently addressed in the context of spintronics' nanowires \cite{lecomte_2009_PhysRevB_80_054413}.

\section{Observables as probe of disorder: statics \textit{versus} dynamics}
\label{section-observables}

There are essentially two basic questions that can be addressed regarding DES and their experimental realizations: how do they look like and how do they respond when one pulls on them.
By comparing the theoretical and numerical predictions of generic DES to measurements on experimental setups, it is possible on one hand to test the adequacy of a DES modelling, and on the other hand to identify the universality class of dimensionality, elasticity and disorder of a particular physical realization, and then to extrapolate to its other possible disorder-conditioned features.

From an analytical point of view, the competition between elasticity, disorder and thermal/quantum fluctuations is treated via two statistical averages of an observable $\mathcal{O}$, respectively the thermal/quantum average ${\moy{\mathcal{O}}}$ and the disorder average ${\overline{\mathcal{O}}}$.
The two underlying assumptions to this procedure are the \textit{ergodicity} and a \textit{self-averaging disorder}, which imply essentially that the computed quantity $\overline{\moy{\mathcal{O}}}$
should match the measurement of  the observable $\mathcal{O}$ on an equilibrated and sufficiently large experimental sample (here with quenched disorder).

Thereafter we discuss briefly the statics and the dynamics of DES, and then we mention the main methods used to tackle them. For short pedagogical reviews on the subject and further references, see \cite{giamarchi_2006_arXiv:0503437,TG_DES_Springer}.

\subsection{Statics: geometrical fluctuations and roughness}
\label{section-statics}

In statics the main information that can be accessed experimentally is the configuration of the system, described by the displacement field ${\vec{u}_{\vec{z}}}$ defined with respect to a given reference configuration (which is somehow arbitrarily chosen) as in Fig.\ref{fig-def-uz-dm1}.

\begin{figure}
\includegraphics[width=8cm]{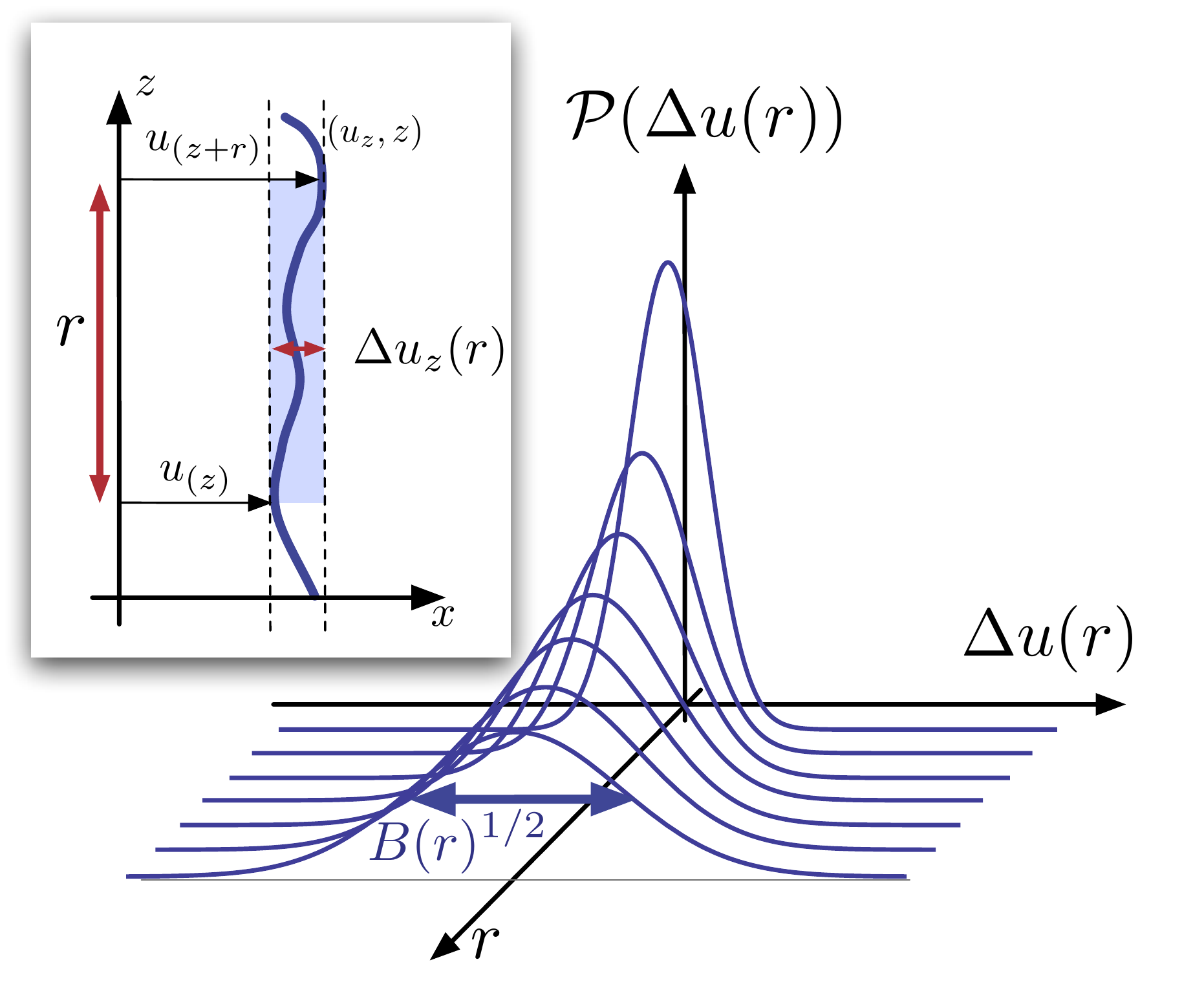}
\caption{Definition of the relative displacements ${\Delta \vec{u}_{\vec{z}} (\vec{r})}$ for a 1D interface and their translational-invariant PDF ${\mathcal{P}(\Delta u(\vec{r}))}$ after the disorder average. If ${\mathcal{P}(\Delta u(\vec{r}))}$ is Gaussian, its main feature is its variance ${\overline{\moy{\Delta \vec{u} (\vec{r})^2}}=B(\vec{r})}$, namely the interface roughness as a function of the lengthscale.}
\label{fig-relative-displacements}
\end{figure}

The geometrical fluctuations of ${\vec{u}_{\vec{z}}}$ can be quantitatively characterized by the probability distribution function (PDF) of the relative displacements ${\Delta \vec{u}_{\vec{z}}(\vec{r})} {\equiv \vec{u}_{\vec{z}+\vec{r}} - \vec{u}_{\vec{z}}}$ at a given lengthscale $\vec{r}$.
Assuming rightfully a translational invariance after disorder-averaging, this PDF ${\mathcal{P}(\Delta u(\vec{r}))}$ can fairly enough be approximated by a widening Gaussian \cite{rosso_2005_JStatMech2005_L08001} as illustrated in Fig.\ref{fig-relative-displacements},
and its main feature is thus its variance, namely the \textit{roughness function} ${B(\vec{r})
 \equiv \overline{\moy{\Delta \vec{u} (\vec{r})^2}}}$ or its corresponding \textit{structure factor} ${S(\vec{q})}$:
\begin{equation}
 B(\vec{r})
 = \int_{\mathbb{R}^d} \frac{d \vec{q}}{(2 \pi)^d} 2\argc{1- \cos(\vec{q} \cdot \vec{r})} S(\vec{q})
\end{equation}
Those quantities are thus the mere two-points correlation functions of ${\vec{u}_{\vec{z}}}$ and its Fourier transform, for which we are hopefully well-equipped for an analytical treatment.
Moreover, if the system displays a scale invariance on a whole range of lengthscales, it is expected to behave logarithmically or rather as a power law ${B(r) \sim r^{2\zeta}}$ or ${S(q) \sim q^{-d-2\zeta}}$ defining the \textit{roughness exponent} $\zeta$.

For periodic systems, the displacement field is again defined with respect to a reference `equilibrium' configuration, this time a lattice of reciprocal vector $\vec{K}_0$. The analogous quantity to the roughness is then the  \textit{translational order correlation function} ${C_{\vec{K}_0}(\vec{r})}$ defined as:
\begin{equation}
 C_{\vec{K}_0} (\vec{r})
 \equiv \overline{\moy{e^{i \vec{K}_0 \argp{\vec{u}_{\vec{r}} - \vec{u}_{\vec{0}}}}}}
\end{equation}
and its Fourier transform the structure factor ${S(\vec{K}_0 + \vec{q})}$.
If the theory is close to a Gaussian as in \cite{giamarchi_ledoussal_1995_PhysRevB52_1242}, it is actually simply related to the roughness since we have then ${C_{\vec{K}_0} (\vec{r})=\exp \argc{-\frac{\vec{K}_0^2}{2}B(\vec{r})}}$.

In practice, in order to measure the roughness and its related quantities, one has first to take a picture (a `snapshot') of the system, then chose a reference configuration (typically flat for an interface or a perfect lattice for periodic systems) and define the corresponding displacement field $\vec{u}_{\vec{z}}$, and finally compute the appropriate functions $B(\vec{r})$, ${C(\vec{r})}$ or  $S(\vec{q})$.
In a logarithmic representation, the roughness usually exhibits a power-law (or logarithmic) behavior ${B(r)\sim A(c,D,T,\xi) \cdot r^{2\zeta}}$ up to a certain lengthscale above which it saturates;
the main focus then is on the value(s) of the corresponding roughness exponent(s) ${\zeta}$, as a signature of dominant physics at a given lengthscale range depending on the universality class to which the system belongs.
There are however other important features also at our disposal to probe the disorder-conditioned properties of the system, such as the power-laws prefactors and their possible temperature-dependence e.g. ${A(c,D,T,\xi) \sim T^{2\text{\thorn}}}$ (with the \textit{thorn} exponent \thorn), the crossover lengthscales including the saturation lengthscale itself and last but not least the possible non-Gaussianity of the PDF ${\mathcal{P} (\Delta u (\vec{r}))}$ (that could then be due to finite-statistics artefacts or to a physical origin) \cite{santucci_2007_PhysRevE75_016104}.
The special case of a 1D interface with a short-ranged elasticity and a RB disorder is discussed at length in the section \ref{section-1Dinterface}.

Such an analysis is of course also possible in order to study the dynamical geometrical fluctuations of the system, but that requires a time-resolved measurement technique, and the possibility to take an actual snapshot of the system, which is not always the case (e.g. for surface-scanning techniques such as the Scanning Tunnelling Microscopy or Atomic Force Microscopy).

\subsection{Dynamics: velocity-force characteristic}
\label{section-dynamics}

The knowledge of the statics of a disordered system is not a sufficient criterion to decide on its glassiness, since it may look quasi-ordered despite the disorder (as e.g. periodic DES displaying power-law decaying Bragg peaks).
Because of its disorder-induced metastability, it may moreover be difficult to even discriminate equilibrium from out-of-equilibrium configurations, so it is necessary to study specifically the dynamics of the system.

The characterization of the response of a disordered system under an external force is fundamental for applications, and is thus widely studied in experimental setups.
Applying an external force on a DES, such as an external magnetic field on a ferromagnetic domain wall or an electrical current on a vortex lattice in a type-II superconductor, it is indeed possible to set those systems into motion, which is highly non-trivial since they are after all emergent structures defined by a shift of the order parameter.
For example, an external magnetic field favors a given magnetization direction, makes the favored ferromagnetic domains grow and thus displace the boundaries between the different phases, thus the domain walls effectively move.

Ideally we would like to have access to the whole time-resolved displacement field ${\vec{u}_{\vec{z}}(t)}$, as in numerical simulations but also in experiments such as e.g. the imaging of the imbibition line of a fluid on a disordered substrate \cite{santucci_2011_EurPhysLett94_46005},
of a crack front along an heterogeneous weak plane \cite{tallakstad-santucci_2011_PhysRevE83_046108} by ultra-fast CCD camera,
or of avalanches in ferromagnetic thin films \cite{repain_2004_EurPhysLett68_460}.
The quantities of interest, beyond the displacement field itself ${\vec{u}_{\vec{z}}(t)}$,
are usually the position of its center of mass ${\vec{u}_{\text{CM}}(t)}$,
its velocity ${v(t)= \partial_t \vec{u}_{\text{CM}}(t)}$ and the fluctuations of the velocity ${\partial_t \vec{u}_{\vec{z}}(t)}$ itself.
However, for a complete characterization, the geometrical fluctuations around the center of mass ${\delta \vec{u}_{\vec{z}}(t) = \vec{u}_{\vec{z}}(t) - \vec{u}_{\text{CM}}(t)}$ should also be addressed,
with the PDF of the \textit{ad hoc} dynamical relative displacements at fixed time ${\Delta \vec{u}_{\vec{z}}(\vec{r}) \equiv \delta \vec{u}_{\vec{z}+ \vec{r}}(t) - \delta \vec{u}_{\vec{z}}(t)}$ and its corresponding variance ${B(\vec{r},t)}$, structure factor ${S(\vec{q},t)}$ or higher moments similarly to the statics.
\begin{figure}
\includegraphics[width=9cm]{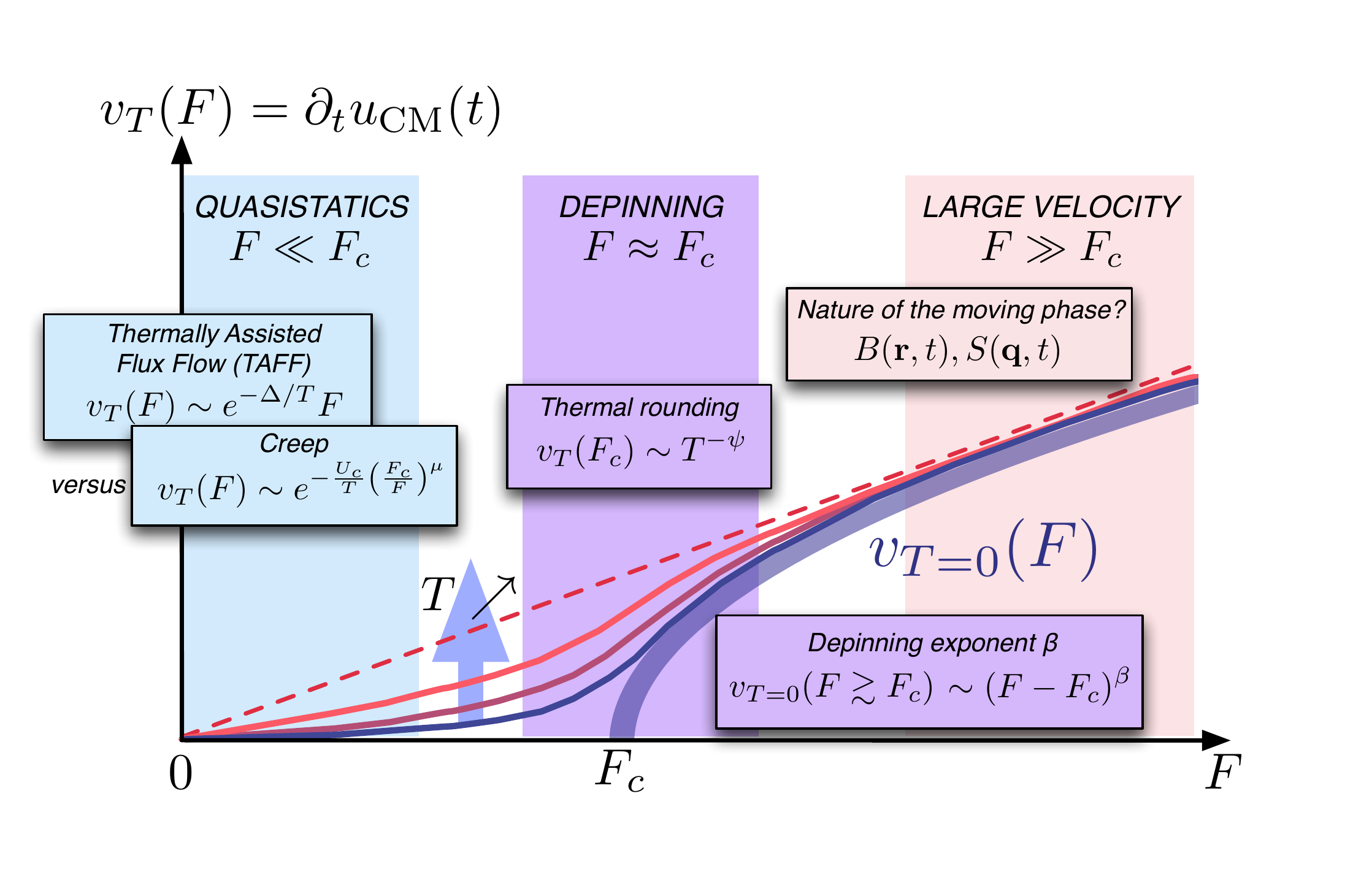}
\caption{Velocity-force characteristics: steady-state velocity of the center-of-mass position $v_{T}$ as a function of a constant external force $F$. The large arrow shows the displacements of the curves ${v_T(F)}$ with an increasing temperature $T$, coupled with a gradation from blue to red curves.}
\label{fig-velocity-force}
\end{figure}

Analytically, numerically and experimentally there has mainly been a focus on the determination of the steady-state velocity of the center-of-mass under a constant driving force $F$, with the temperature-dependent velocity-force characteristic $v_T(F)$ illustrated schematically in Fig.\ref{fig-velocity-force}.
At zero temperature, the picture is superficially similar to that of a critical phenomenon, with the steady-state velocity $v_T$ as an order parameter, the driving force $F$ as the controlling parameter and the disorder-dependent critical force $F_c$.
Below $F_c$ the DES might relax from a given configuration with a transient velocity $v(t)$ \cite{kolton-rosso_2006_PhysRevB74_140201}, but then it is stuck so ${v_{T=0}(F<F_c)=0}$.
Above $F_c$ it acquires a finite steady-state velocity ${v_{T=0}(F>F_c)>0}$ through the disordered energy landscape, which actually takes off with a critical \textit{depinning} exponent ${v_{T=0} (F \gtrsim F_c) \sim (F-F_c)^{\beta}}$.
With a finite temperature, there are thus three regimes of interest for the thermal effects on the dynamics:
first the \textit{depinning} regime at ${F \approx F_c}$ with a possible thermal-rounding power law $v_T(F_c)\sim T^{-\psi}$ \cite{bustingorry_2008_epl-81-2-26005,bustingorry_2009_PhysicaB404_444};
secondly the \textit{large velocity} regime at ${F \gg F_c}$ where $v_T \approx \eta F$ which defines the friction coefficient $\eta$ of the DES, and the possible characterization of the nature of this moving phase via its dynamical geometrical fluctuations;
and finally the \textit{creep} regime at ${F \ll F_c}$ where thermal fluctuations allow the system to explore the disordered energy landscape and overcome some of its barriers.

If there was only one characteristic energy barrier $\Delta$, the response of the system to a very small driving force would be linear ${v_T(F) \sim e^{-\Delta/T} F}$
(`Thermally Assisted Flux Flow').
However such DES actually exhibits a disorder-induced metastability and thus there is no fixed characteristic energy barrier, and the response is rather a stretched exponential:
\begin{equation}
 v_T(F) \propto \exp \argc{- \frac{U_c}{T} \argp{\frac{F_c}{F}}^{\mu}}
 \label{equa-creepformula}
\end{equation}
This phenomenological `creep' formula can be related to the static quantities assuming that the creepy interface moves forward with a succession of avalanches of typical size $L_c$ (`Larkin domains').
Assuming that $L_c$ would be the characteristic crossover lengthscale in the static roughness, and comparing by scaling the elastic, disorder and external force energies at stake at this lengthscale \cite{nattermann_1987_EPL4_1241,ioffe_vinokur_1987_JPhysC20_6149,chauve_2000_ThesePC_PhysRevB62_6241,giamarchi_2006_arXiv:0503437},
the creep exponent is predicted to be ${\mu = (d-2+2 \zeta_{RM})/(2-\zeta_{RM})}$, with $d$ the dimension and $\zeta_{RM}$ the static asymptotic roughness exponent.
This behavior has been spectacularly pointed out for the 1D interface on several order of magnitudes in ferromagnetic DWs \cite{lemerle_1998_PhysRevLett80_849} $ \mu = 0.24 \pm 0.04$, $ \zeta = 0.69 \pm 0.07$, $ d=1$ and in numerical studies \cite{kolton_2009_PhysRevB79_184207}.
However the content of the typical energy $U_c$ and depinning force $F_c$ is still an open question.
The single-crossover scenario predicts ${U_c=c \xi^2 /L_c}$ the typical energy barrier at the lengthscale $L_c$ and ${F_c=c \xi / L_c^2}$ the depinning force of a Larkin domain, with $\xi$ the width of the interface, but this prediction has not been reproduced yet in numerics even for the 1D interface.

\subsection{Methods}
\label{section-methods}

In statics, a given configuration with a displacement field ${\vec{u}_{z}}$ is weighted by a Boltzmann weight ${\propto e^{-\mathcal{H}\argc{\vec{u},V}/T }}$ (with the Boltzmann constant $k_B=1$ so that the temperature $T$ has actually the units of an energy),
whereas in dynamics the Hamiltonian ${\mathcal{H} \argc{u,V}}$ is replaced by a Martin-Siggia-Rose action constructed from a Langevin equation for the time-evolution of the displacement field ${\vec{u}_{\vec{z}}(t)}$.
The statistical averages $\overline{\left< \mathcal{O} \right>}$ for the observables $\mathcal{O}$ listed in the previous sections are then computed analytically with respect to those statistical weights.

Pure scaling arguments on the statistical averages of observables, via the scaling of the Hamiltonian, of the disorder distribution, of the Langevin equation or of the action, can yield very powerful predictions yet they have to be interpreted carefully, and can often provide a rather \textit{a posteriori} short explanation of otherwise painfully computed results.
For example, the Flory or Imry-Ma `mean-field' argument
provides a value $\zeta_{\text{F}}$ quite close to the exact physical roughness exponent $\zeta$.

The two main analytical tools used on DES are
on one hand the \textit{Functional Renormalization Group} (FRG) where the whole disorder correlator \eqref{equa-gendiscorr} evolves under the renormalization procedure
\cite{fisher_1986_PhysRevLett56_1964,giamarchi_ledoussal_1995_PhysRevB52_1242,ledoussal-wiese-chauve_2004_PhysRevE69_026112,wiese-ledoussal_2007_arXiv:cond-mat/0611346,ledoussal_2008_arXiv:0809.1192},
and on the other hand the \textit{Gaussian Variational Method} (GVM) as introduced by M\'ezard and Parisi on DES, and involving \textit{Replica} to treat the disorder \cite{castellani-cavagna_2005_JStatMechP05012,book_beyond-MezardParisi}.

Numerically there are in particular very efficient algorithm in 1D in order to address the dynamics and the static of the 1D interface, starting from a \textit{Langevin equation} both at zero \cite{rosso_krauth_2002_PhysRevE65_025101} and at small but finite temperature \cite{kolton-rosso_2006_PhysRevLett97_057001,kolton_2009_PhysRevB79_184207}.

\section{Static roughness of a 1D interface at finite temperature}
\label{section-1Dinterface}

In this section, we focus on the particular case of the 1D interface in quenched RB disorder, as an illustration of a DES modelling, of the computation and interpretation of the static roughness function via GVM and scaling arguments, and of the link to the quasistatic creep regime.

The study of the 1D interface is at the crossroad between
a fundamental interest in the peculiarities of low-dimensional systems,
several mappings on related statistical-physics problems in the Kardar-Parisi-Zhang (KPZ) universality class (including the 1+1 Directed Polymer (DP)) \cite{corwin_2011_arXiv:1106.1596,krug_1997_AdvPhys_46_139,halpin_zhang_1995_PhysRep254},
and experimental realizations of effective 1D interfaces such as ferroic or ferromagnetic domain walls (DW) in thin films.

A first example of such experiments are DWs in ferroelectric thin films of Pb(Zr$_{0.2}$Ti$_{0.8}$)O$_3$ which display out-of-plane polarized domains, written and probed by atomic-force-microscopy technique \cite{tybell_2002_PhysRevLett89_097601,paruch_2005_PhysRevLett94_197601}.
This imaging technique probes by definition the surface of the film, which is however sufficiently thin (typically 50-60 nanometers) so that the measured DWs are effective 1D interfaces convoluted on the whole thickness of the film.
A second example, which is believed to be a prototype of the DES model presented thereafter, are ferromagnetic DWs in ultrathin films of Pt/Co/Pt (just a few atomic layers) with out-of-plane magnetization domains \cite{lemerle_1998_PhysRevLett80_849,repain_2004_EurPhysLett68_460,metaxas_2007_PhysRevLett99_217208}.

The internal structure of those DWs have not been accessed, since their typical width (respectively a few unit crystal cells and $\sim 10$ nm) is still below the resolution (respectively nanometric and micrometric), along with their transverse geometrical fluctuations at comparable small lengthscales.
However, even though their static roughness cannot be observed at sufficiently small lengthscales yet, it could have indirect consequences on their quasistatic dynamics.
Indeed, in order to obtain the phenomenological creep formula (cf. section \ref{section-dynamics}),
the main assumption is that there is a \textit{single} characteristic lengthscale in the statics (the Larkin length $L_c$) which coincides with the typical size of avalanches in the quasistatic dynamics.
The scaling prediction for the creep exponent worked surprisingly well in ferromagnetic DWs \cite{lemerle_1998_PhysRevLett80_849} for the roughness and creep exponents $\zeta$ and $\mu$ (cf. section \ref{section-dynamics}),
however the discrepancies regarding the temperature-dependence of the typical energy barrier $U_c$ observed in numerical results \cite{kolton_2005_PhysRevLett94_047002} suggested to reexamine the low-temperature static roughness at small lengthscales.

\subsection{Full DES model of the 1D interface}

Following the DES recipe of the section \ref{section-DESrecipe},
we consider a 1D interface (${d=m=1}$) embedded in a continuous and infinite space ${(z,x) \in \mathbb{R}^2}$, with implicit ultra-violet and infra-red cutoffs ultimately irrelevant in our computations.
Restricting ourselves to the case without overhangs nor bubbles,
each configuration is characterized by a univalued displacement field ${u_z}$ with respect to a flat configuration defined by the $z$ axis.

We assume a short-range elasticity in the elastic limit as in \eqref{equa-Hel} and a quenched RB weak disorder uncorrelated along its internal direction $z$ as in \eqref{equa-Hdis}.
This very particular form of elasticity is central for the mappings to other statistical physics problems and for the \textit{statistical tilt symmetry} (STS), a fundamental property for 1D interfaces in presence of disorder \cite{schulz_1988_JStatPhys51_1,hwa_1994_PhysRevB49_3136,fisher_huse_1991_PhysRevB43_10728}.

\begin{figure}
\includegraphics[width=8cm]{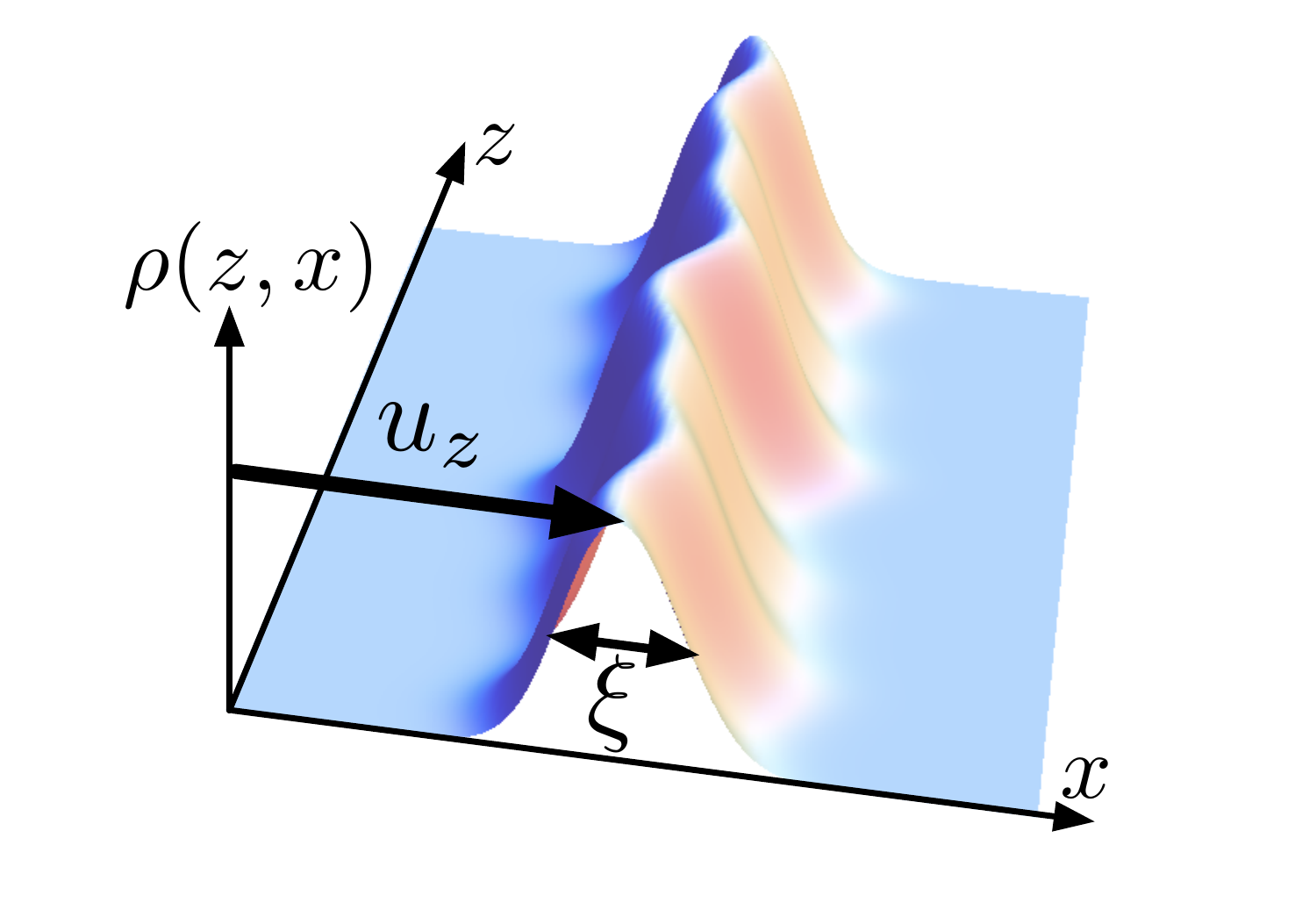}
\caption{Gaussian profile of a 1D interface, described by the normalized density ${\rho_{\xi} (z,x)}$ as a Gaussian of variance $\xi$ centered at fixed $z$ on the interface mean position $u_z$.}
\label{fig-profile3DGauss}
\end{figure}

The disorder correlator \eqref{equa-gendiscorr} is finally taken as a Gaussian of variance ${\xi_x^2= 2 \xi^2}$ and amplitude $D$:
\begin{equation}
 R_{\xi_x}(x)
 = D \cdot \frac{e^{-x^2/(4 \xi^2)}}{\sqrt{4\pi} \xi}
 = D \int_{\mathbb{R}} \frac{d \lambda}{2\pi} \cdot e^{i \lambda x} e^{-\lambda^2 \xi^2}
 \label{equa-Gaussdiscorr}
\end{equation}
So there are two distinct Gaussian hypothesis regarding the disorder: on one hand its distribution ${\mathcal{P}\argc{V}}$ is assumed to be Gaussian and thus fully described by its disorder correlator (a physical assumption in the weak disorder limit), and on the other hand its disorder correlator itself is chosen to be a Gaussian.
In this choice ${R(x)}$ decreases sufficiently fast to be random-bond and it encodes explicitly a finite disorder correlation length $\xi$ in a convenient way for computations.
In an alternative point of view, this parameter $\xi$ corresponds to a finite width of the 1D interface, whose spatial extension in the $x$ direction is described by a Gaussian density ${\rho_{\xi}(z,x)}$ of variance $\xi^2$ centered on ${(z,u_z)}$ as illustrated in Fig.\ref{fig-profile3DGauss}; if the random-potential is $\delta-$correlated in both ${(z,x)}$ directions, the effective disorder ${\int_{\mathbb{R}} dx \cdot \rho_{\xi} (z,x) V(z,x)}$ has precisely the correlator \eqref{equa-Gaussdiscorr}.

\subsection{DES characteristic scales and scaling arguments} \label{section-scaling}

In this minimal DES model, there are thus four characteristic scales:
the elastic constant $c$ which gives the energy per unit of length of the interface and essentially fixes an energy-scale reference;
the strength of disorder $D$ or the amplitude of the disorder correlator;
the finite temperature $T$ which quantifies the energy of the surrounding thermal bath;
and the typical width or disorder correlation length $\xi>0$ or the variance of the disorder correlator.

The main consequence of the interplay between a finite ${T>0}$ and a finite ${\xi>0}$ on the 1D interface geometrical fluctuations is the existence of two regimes in temperature for its static roughness.
Those two regimes are separated by the single characteristic temperature ${T_c=(\xi c D)^{1/3}}$ which naturally involves all the scales at stake in the DES model.
Physically this can actually be understood by a comparison between the microscopic width $\xi$ and an effective width ${\xi_{\text{th}}}$ due to thermal fluctuations.
At high temperature (${T>T_c}$) the thermal fluctuations erase the existence of the microscopic width (${\xi<\xi_{\text{th}}}$), so with the only two remaining scales $\arga{D,T}$ one can define a single characteristic lengthscale $r_*$, presumably the unique crossover between two roughness regimes (purely thermal at small-lengthscales and dominated by disorder at large lengthscales).
On the contrary, at low temperature (${T<T_c}$) ${\xi>\xi_{\text{th}}}$ so the microscopic width matters and there is a whole intermediate roughness regime which leads to the definition of at least two characteristic crossover lengthscales $r_1$ and $r_0$, the latter marking the entrance in the asymptotic random-manifold regime dominated by disorder.

The bare scaling of those different quantities $r_*$, $\xi_{\text{th}}$, $T_c$ and $r_0$ can actually be extracted from simple scaling arguments, by rescaling the spatial coordinates ${z=b\bar{z}}$ and ${x=a\bar{x}}$ in the whole roughness function:
\begin{equation}
\begin{split}
 B(r;c,D,T,\xi)
 &= \int \mathcal{D}V \, \mathcal{P} \argc{V} \moy{(u_{z+r}-u_{z})^2}_V \\
 \moy{(u_{z+r}-u_{z})^2}_V
 &= \frac{\int \mathcal{D}u \cdot (u_{z+r}-u_{z})^2 \cdot e^{-\mathcal{H}\argc{u,V}/T}}{\int \mathcal{D}u \cdot e^{-\mathcal{H}\argc{u,V}/T}}
\end{split}
\end{equation}
Indeed we have to assume first that the random potential $V$ `scales in distribution' consistently with respect to the scaling of its disorder correlator, leading thus to
$${\mathcal{H}_{\text{dis}} \argc{u_z,V} \vert_{D,\xi} \stackrel{d}{=} \argp{\frac{bD}{a}}^{1/2} \mathcal{H}_{\text{dis}} \argc{u_{(z/b)}/a,V} \vert_{D=\xi=1}}$$
Since ${\mathcal{H}_{\text{el}} \argc{u_z}\vert_{c} = \frac{c a^2}{b} \mathcal{H}_{\text{el}} \argc{u_z}\vert_{c=1}}$,
if we assume that the whole Hamiltonian ${\mathcal{H}=\mathcal{H}_{\text{el}}+\mathcal{H}_{\text{dis}}}$ can be rescaled with an overall prefactor, a same scaling for the two parts of the Hamiltonian imposes the relation ${a=D^{1/2} c^{-1} b^{\zeta_F}}$ with the Flory exponent ${\zeta_F=3/5}$ for ${d=m=1}$.
Finally we can either absorb all the temperature dependence into an effective thermal width (fixing the temperature, $\xi$ can still be properly neglected so that corresponds to the high-$T$ regime):
\begin{equation}
 \begin{split}
  B(r;c,D,T,\xi) = \xi_{\text{th}}^2 &\, B (r/r_*;1,1,T=1,\xi/\xi_{\text{th}}) \\
 \xi_{\text{th}}(T)=\frac{T^3}{cD} \, ,& \quad r_*(T)=\frac{T^5}{cD^2}
 \end{split}
 \label{equa-scalinghighT}
\end{equation}
or include all the width dependence into a characteristic temperature (fixing the width, the temperature can be pushed arbitrarily low, so that corresponds to the low-$T$ regime):
\begin{equation}
 \begin{split}
 B(r;c,D,T,\xi) = \xi^2 &\, B (r/r_0;1,1,T/T_c,\xi=1) \\
 T_c(\xi) = (\xi c D)^{1/3} \, ,& \quad r_0(\xi) = \xi^{5/3} c^{2/3} D^{-1/3}
 \end{split}
 \label{equa-scalinglowT}
\end{equation}
Those two raw scalings connect of course, via the conditions ${\xi_{\text{th}}(T_c)=\xi}$ and ${r_*(T_c)=r_0(\xi)}$.

To give a physical meaning to the different quantities defined by these scaling arguments, one needs to make additional non-trivial assumptions, essentially on the behavior of the scaling functions ${B(\bar{r},1,1,1,\bar{\xi})}$ and ${B(\bar{r},1,1,\bar{T},1)}$, e.g. a power-law behavior ${\sim \bar{r}^{2\zeta}}$.
For example the Flory exponent ${\zeta_F}$ could be interpreted as the physical roughness exponent ${\zeta}$ if there was only one single power-law behavior in the roughness, thus the whole system would display a true scale invariance at all lengthscales and its Hamiltonian could indeed be rescaled with a scaling factor ${a \sim b^{\zeta_F}}$.
However there are at least two different roughness power-law regimes, at small lengthscales a thermal behavior with ${B_{\text{th}}(r) = Tr/c \sim (T^{1/2}r^{1/2})^2}$ and at large lengthscales the asymptotic RM regime with ${B_{\text{asympt}}(r)\sim (T^{\text{\thorn}}r^{2/3})^2}$.
Even though the asymptotic roughness exponent of exact value ${\zeta_{RM}=2/3}$ can be shown to be robust with respect to the addition of a finite width ${\xi>0}$ \cite{agoritsas_2011_FHHprov}, its temperature-dependence (and \textit{thorn} exponent $\text{\thorn}$) could \textit{a priori} be modified, as emphasized by the two opposite FRG regimes of zero-temperature fixed-point \cite{chauve_2000_ThesePC_PhysRevB62_6241} \textit{versus} high-temperature \cite{bustingorry_2010_PhysRevB82_140201}.

\subsection{GVM roughness of the 1D interface}

In order to go beyond those scaling predictions and to have an explicit roughness function ${B(r;c,D,T,\xi)}$ connecting all the lengthscales at ${\xi>0}$ and ${T>0}$,
we have computed it in full details in the reference \cite{agoritsas_2010_PhysRevB_82_184207} in a Gaussian Variational Method (GVM), compatible with the fairly Gaussian distribution ${\mathcal{P}(\Delta u(r))}$ observed in numerics \cite{halpin_1991_PhysRevA44_R3415}.
This scheme has already been applied to DES, both in periodic systems \cite{giamarchi_ledoussal_1995_PhysRevB52_1242} as well as manifolds \cite{mezard_parisi_1991_replica_JournPhysI1_809}, in the latter the disorder correlator \eqref{equa-Gaussdiscorr} was assumed to decay in a power law, whereas we focused in \cite{agoritsas_2010_PhysRevB_82_184207}  on the consequences of its finite variance ${\sim \xi^2}$.

We first used the so-called  `replica trick', well-known in the study of spin glasses \cite{book_beyond-MezardParisi}, in order to average first over the disorder and so to transform the random part ${\mathcal{H}_{\text{dis}}}$ in the full Hamiltonian of one interface (${u_1}$), into an effective non-random coupling between $n$ copies of the interface (${\vec{u}=\arga{u_1,\dots,u_n}}$), in the overall peculiar limit of ${n \to 0}$.
The effective `replicated' Hamiltonian ${\widetilde{\mathcal{H}}\argc{\vec{u}}}$ thus obtained was then approximated in the statistical thermal average via GVM, i.e. it was replaced by a \textit{quadratic} replicated Hamiltonian ${\mathcal{H}_0 \argc{\vec{u}}}$ optimized by minimizing its corresponding variational free energy.
Note that the GVM approximation is the third Gaussian hypothesis in this study of the 1D interface, based on numerical results \cite{halpin_1991_PhysRevA44_R3415,rosso_2005_JStatMech2005_L08001}.

\begin{figure}
\includegraphics[width=8cm]{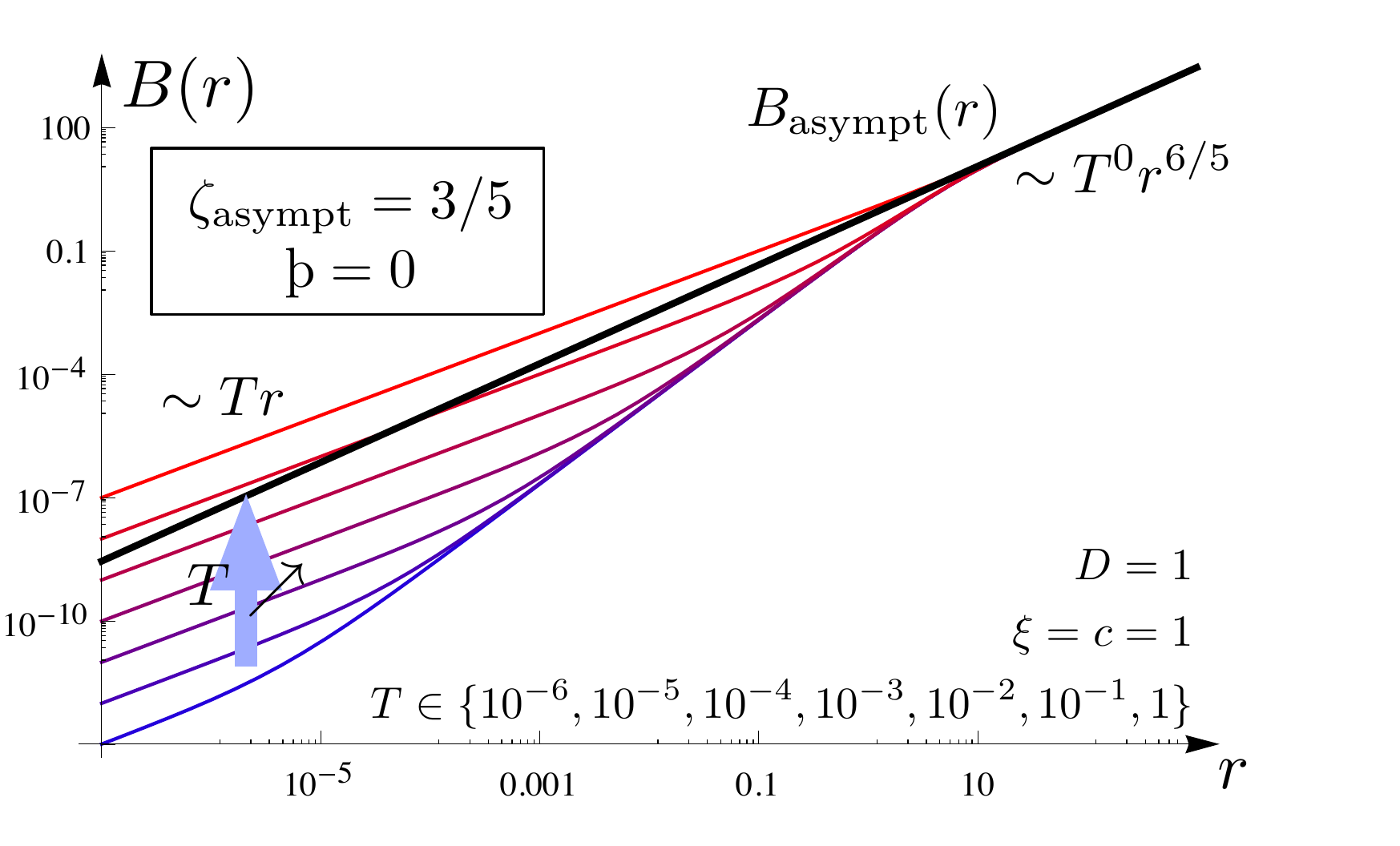}
\caption{Roughness function $B(r)$ at fixed disorder, obtained by GVM on the full DES model of a 1D interface. The large arrow shows the displacements of the curves with an increasing temperature $T$, coupled with a gradation from blue to red curves}
\label{fig-4b-PRB2010}
\end{figure}

In the replica formulation, the Ansatz of the GVM self-energy ${\argc{\sigma}(u)}$ (${u \in \argc{0,1}}$) was taken to be full-replica-symmetry-breaking (full-RSB) below a single cutoff ${v_c\in \argc{0,1}}$ and replica-symmetric (RS) above $v_c$;
the full-RSB encodes physically the existence of metastability  at large lengthscales (via a continuum of self-energies) and the RS plateau the thermal fluctuations of the interface at small lengthscales (via a unique typical self-energy).
This leads to the following roughness function as a function of the lengthscale $r$, illustrated in Fig.\ref{fig-4b-PRB2010}:
\begin{eqnarray}
 B(r)
 &=& \frac{Tr_0}{c} \argp{\frac{r}{r_0} + \bar{B}_{\text{dis}}\argp{\frac{r}{r_0}}} \label{equa-1D-roughness-1} \\
 \bar{B}_{\text{dis}}(\bar{r})
 &=& \frac{1}{v_c} \sum_{k=2}^{\infty} \frac{(-\bar{r})^k}{k!} \argp{\frac{1}{5k-6}+(1-v_c)} \label{equa-1D-roughness-2} \\
 r_0 &=& \frac{5^5 \pi}{3^7} \frac{1}{cD^2} \argp{\frac{T}{v_c}}^5 \label{equa-1D-roughness-3} \\
 v_c^6 &=& \widetilde{A}_1 (5/6 - v_c) \, , \quad \widetilde{A}_1=\frac{5^5 \pi}{2 \times 3^7} \argp{\frac{T}{T_c}}^6 \label{equa-1D-roughness-4}
\end{eqnarray}
All the $\xi$-dependence is actually contained in the full-RSB cutoff $v_c(\xi)$ since ${T_c \equiv (\xi c D)^{1/3}}$.
At low temperature it grows linearly in temperature ${v_c \sim T/T_c}$ leading to the same scaling prediction ${r_0 \sim \xi^{5/3}c^{2/3} D^{-1/3}}$ as in \eqref{equa-scalinglowT}, whereas at high temperature it saturates at ${v_c \lesssim 5/6}$ and we recover ${r_0\sim T^5/(cD^2) = r_*(T)}$ as in \eqref{equa-scalinghighT}.

Two artefacts of the GVM computation can be directly seen in Fig.\ref{fig-4b-PRB2010}:
the asymptotic roughness functions at large lengthscales collapse on a single curve ${B_{\text{asympt}}(r) \sim T^0 r^{6/5}}$ i.e. ${\zeta_{\text{asympt}}=3/5=\zeta_{\text{F}}^{\text{1D}}}$ and ${\text{\thorn}=0}$.
So the GVM predicts on one hand the Flory exponent $3/5$ instead of the exact RM roughness exponent ${\zeta_{\text{RM}}=2/3}$; and on the other hand temperature-independent fluctuations at large lengthscales in contradiction with the high-$T$ scaling argument \eqref{equa-scalinghighT} (except if we assume temperature-dependent microscopic parameters ${\arga{D,c,\xi}}$).

\subsection{Effective DP toy model and its GVM roughness}

An alternative to this GVM computation on a full DES model is to use the exact mapping of the 1D interface on the 1+1 Directed Polymer \cite{mezard_parisi_1992_JPhysI02_2231}, in order to construct an effective `toy model' \textit{containing additional physical ingredients in a controlled way at a fixed lengthscale}, and only then to perform the GVM approximation for the roughness.

Assuming as always that there is no overhangs nor bubbles,
a segment of length $r$ of the 1D interface ${(z,u_z) \in \mathbb{R}^2}$ can be seen as a directed polymer with one extremity fixed at the origin ${(0,0)}$, growing in `time' along a trajectory ${(t',y(t')) \in \mathbb{R}^2}$, until a time $t$ identified with the lengthscale $r$.
With the translations ${t \leftrightarrow r}$ and ${y(t) \leftrightarrow u_z}$, the geometrical fluctuations of the 1D interface at a given lengthscale $r$ are mapped to the DP's endpoint fluctuations at a fixed time $t$,
${\mathcal{P}(\Delta u(r))} \leftrightarrow {\mathcal{P}(y(t))}$
and so their variance define the same roughness function:
\begin{equation}
B(r) \equiv \overline{\moy{\Delta u(r)^2}}
\stackrel{(t\leftrightarrow r)}{\Longleftrightarrow}
\overline{{\moy{y(t)^2}}} \equiv B^{\text{DP}}(t)
\end{equation}
In the absence of disorder, the polymer draws a Brownian random walk and the probability distribution of its endpoint is a pure Gaussian given by
\begin{equation}
 \mathcal{P}_{\text{th}} (y(t))
 = \frac{\exp \argp{-\frac{1}{T}\frac{c y^2}{2t}}}{\sqrt{2 \pi \frac{Tt}{c}}}
 \equiv \exp \argp{-\frac{ F_{\text{th}}(t,y)}{T}}
\end{equation}
where the pseudo-free energy ${F_{\text{th}}(t,y)}$ is the corresponding term to the elastic Hamiltonian \eqref{equa-Hel}.
The RB disorder explored by the DP translates into an effective free energy ${\bar{F}_{\eta}(t,y)}$
such that ${\mathcal{P}(y(t))} {\propto e^{-\argp{F_{\text{th}}(t,y)+\bar{F}_{\eta}(t,y)}/T}}$, and defined as an integrated random phase ${\eta(t,y)}$ in a `random-field' effective formulation:
\begin{equation}
 \bar{F}_{\eta} (t,y) = \argp{\int_y^{\infty} - \int_{-\infty}^y} d \tilde{y} \cdot \eta (t,\tilde{y}) + \text{cte}(t)
\end{equation}
As the corresponding term of the disorder Hamiltonian \eqref{equa-Hdis}, we assumed that at fixed $t$ this effective disorder is Gaussian, i.e. ${\mathcal{P} \argc{\bar{F}_\eta(t)}}$ and ${\mathcal{P} \argc{\eta(t)}}$ are fully defined by their mean value and their two-point correlator.
This is known to be exact in the infinite-time limit, at finite temperature and zero-width \cite{huse_henley_fisher_1985_PhysRevLett55_2924}:
\begin{equation}
\begin{split}
\overline{\eta(\infty,y_1) \eta(\infty,y_2)}
= \widetilde{D} \cdot \delta(y_1-y_2)
\, , \quad \widetilde{D}=\frac{cD}{T} \\
\overline{\argc{\bar{F}_{\eta} (\infty,y_1)-\bar{F}_{\eta} (\infty,y_2)}^2}
= \widetilde{D} \cdot \vert y_1-y_2 \vert
 \end{split}
 \label{equa-FHH-infinite-time}
\end{equation}
where the polymer has completely forgotten its initial condition and displays a translational-invariance invariance in time.
So, based on this infinite-time exact result, we assumed at last that the finite $\xi$ of the microscopic random potential ${V(x,z)}$ translates simply at finite time $t$ into a broadening of the correlator ${\overline{\eta \eta}(t)}$, of variance ${2 \tilde{\xi}^2}$ and generic amplitude $\widetilde{D}$, taken to be Gaussian for the sake of computations and equal to \eqref{equa-Gaussdiscorr} with ${\xi_x=2 \tilde{\xi}^2}$.

\begin{figure}
\includegraphics[width=8cm]{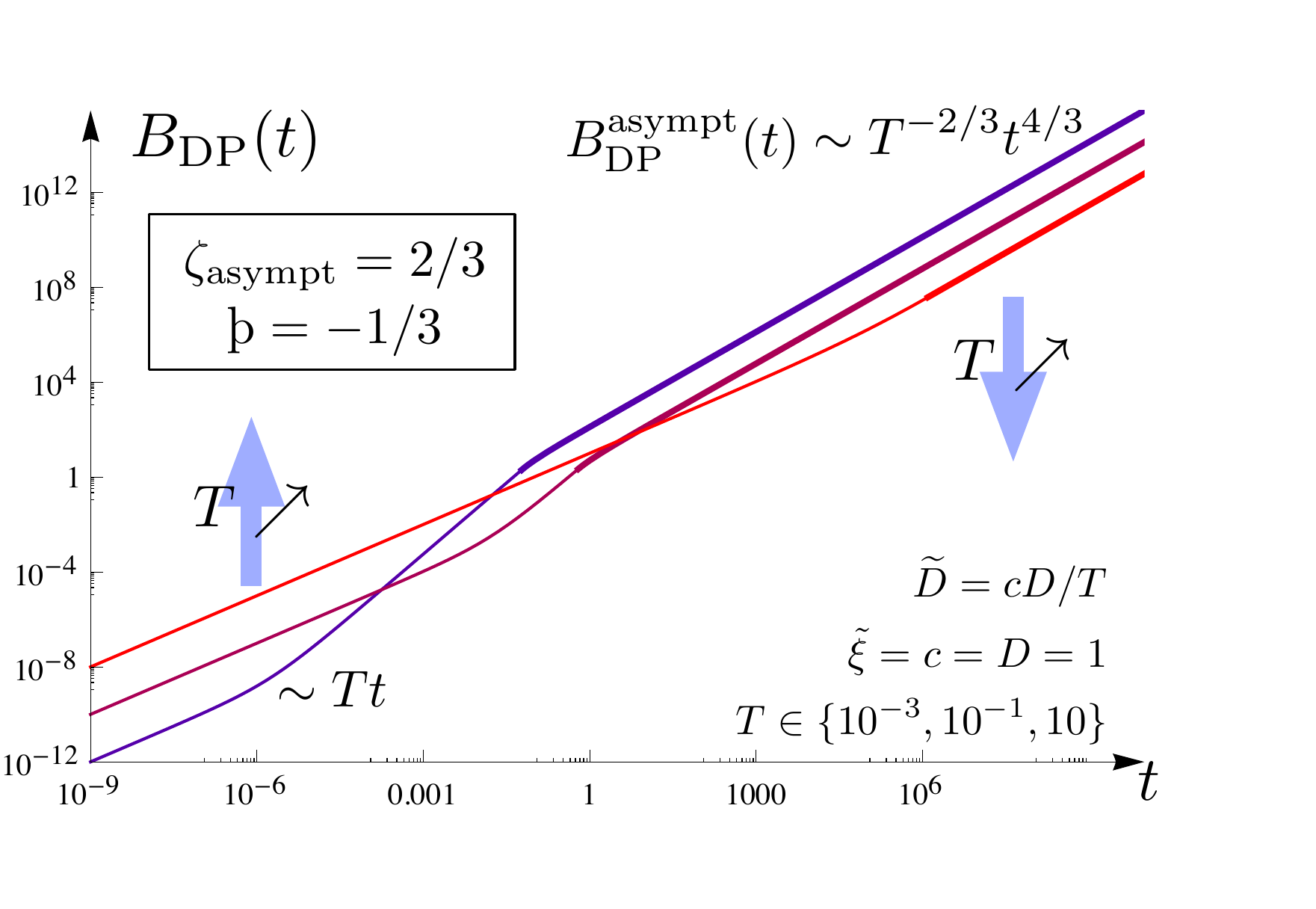}
\caption{Roughness function $B_{\text{DP}}(t)$ assuming a fixed effective disorder ${\widetilde{D}=cD/T}$, obtained by GVM on the Directed Polymer toy model. The large arrows show the displacements of the curves with an increasing temperature $T$, coupled with a gradation from blue to red curves.}
\label{fig-7-PRB2010}
\end{figure}

Performing again a GVM computation of the roughness, with a full-RSB Ansatz for the self-energy below a single cutoff ${u_c \in \argc{0,1}}$ and RS above, we have obtained \cite{agoritsas_2010_PhysRevB_82_184207}, as illustrated in Fig.\ref{fig-7-PRB2010}:
\begin{eqnarray}
\!\!\!\!\!\!\! B_{\text{DP}}(t \geq t_c)
 &=& \frac{3}{2} \argp{\frac{c \widetilde{D}^2}{\pi c^4}}^{1/3} t^{4/3} - \tilde{\xi}^2 \label{equa-DPtoy-roughness-1} \\
\!\!\!\!\!\!\! B_{\text{DP}}(t \leq t_c)
 &=& \frac{Tt}{c} + \frac{\widetilde{D}}{c^2 \sqrt{\pi}} \cdot t^2 \argp{\tilde{\xi}^2 + \frac{Tt}{c}}^{-1/2} \label{equa-DPtoy-roughness-2} \\
 t_c &=& \frac{3^3 \pi}{2^4} \frac{c}{\widetilde{D}^2} \argp{\frac{T}{u_c}}^3 \label{equa-DPtoy-roughness-3} \\
 %
 u_c^4 &=& \widetilde{A}_2 (3/4 - u_c) \, , \quad \widetilde{A}_2=\frac{3^3 \pi}{2^4} \frac{T^4}{(\tilde{\xi} \widetilde{D})^2} \label{equa-DPtoy-roughness-4}
\end{eqnarray}
predictions to compare with \eqref{equa-1D-roughness-1}-\eqref{equa-1D-roughness-4}
with the translation of `time' $t$ into the lengthscale $r$ and the identification of ${t_c}$ with ${r_0}$.

This second GVM procedure is actually carried out \textit{at fixed `time' $t$}, so a priori with `time'-dependent effective parameters ${\tilde{\xi}_t}$ and ${\widetilde{D}_t}$.
We can nevertheless assume that first the disorder correlation length for the DP is essentially the same as for the microscopic random potential i.e. ${\tilde{\xi}=\xi}$, and secondly that $\widetilde{D}$ is given by the constant ${\widetilde{D}=cD/T}$ as in \eqref{equa-FHH-infinite-time}, compatible with the high-temperatures (or zero-width) scaling argument \eqref{equa-scalinghighT}.
This last choice becomes however clearly unphysical in the limit ${T \to 0}$, and in fact a similar scaling argument on the DP toy model suggests that at low temperatures the effective strength of disorder should saturate at ${\widetilde{D}=cD/T_c}$ in order to recover the low-temperatures (or finite-width) predictions \eqref{equa-scalinglowT}, both for ${r_0(\xi)}$ and ${T_c(\xi)}$.
A numerical and analytical study of the DP toy model is currently under preparation \cite{agoritsas_2011_FHHprov}.

With the exact `high-temperature' result ${\widetilde{D}=cD/T}$, all the $\tilde{\xi}$-dependence is contained in the full-RSB cutoff ${u_c(\tilde{\xi})}$ as in \eqref{equa-1D-roughness-4},
and with the two opposite limits ${\widetilde{A}_2 \to 0}$ (${u_c \sim T/(\tilde{\xi}\widetilde{D})^{1/2}}$)  and ${\widetilde{A}_2 \to \infty}$ (${u_c \lesssim 3/4}$) we recover consistently the high-$T$ scaling predictions ${ t_c \sim T^5/(cD^2) = r_*(T)}$ as in \eqref{equa-scalinghighT}.

This second GVM computation on the DP toy model actually gives the same qualitative predictions as for the full 1D-interface model (high- \textit{versus} low-$T$ regimes, intermediate roughness regime at low-$T$), but it yields by construction of the model the correct asymptotic roughness exponent ${\zeta_{\text{asympt}}=2/3=\zeta_{\text{F}}^{toy}}$ and also with ${\widetilde{D}=cD/T}$ a non-zero \textit{thorn} exponent ${\text{\thorn}=-1/3}$ since ${B_{\text{DP}}^{\text{asympt}}\sim T^{-2/3} r^{4/3}}$, compatible with the high-$T$ scaling argument \eqref{equa-scalinghighT}.

\subsection{Larkin length and effective width ${\xi_{\text{eff}}}$}

A physical benchmark for the roughness is the Larkin length ${L_c}$, which is the lengthscale
marking the beginning of the RM asymptotic regime (${ L_c \leftrightarrow r_*,r_0, t_c }$)
and encoded in the GVM approach into the maximum value of the self-energy ${\argc{\sigma}(v_c) \equiv cr_0^{-2}}$.
This length along the \textit{internal} direction of the interface has actually been physically defined by Larkin \cite{Larkin_model_1970-SovPhysJETP31_784} as \textit{the lengthscale at which the typical relative displacement corresponds to the effective width ${\xi_{\text{eff}}}$ of the interface:} ${B(L_c) \sim u(L_c)^2 \sim \xi_{\text{eff}}^2}$.

\begin{figure}
\includegraphics[width=8cm]{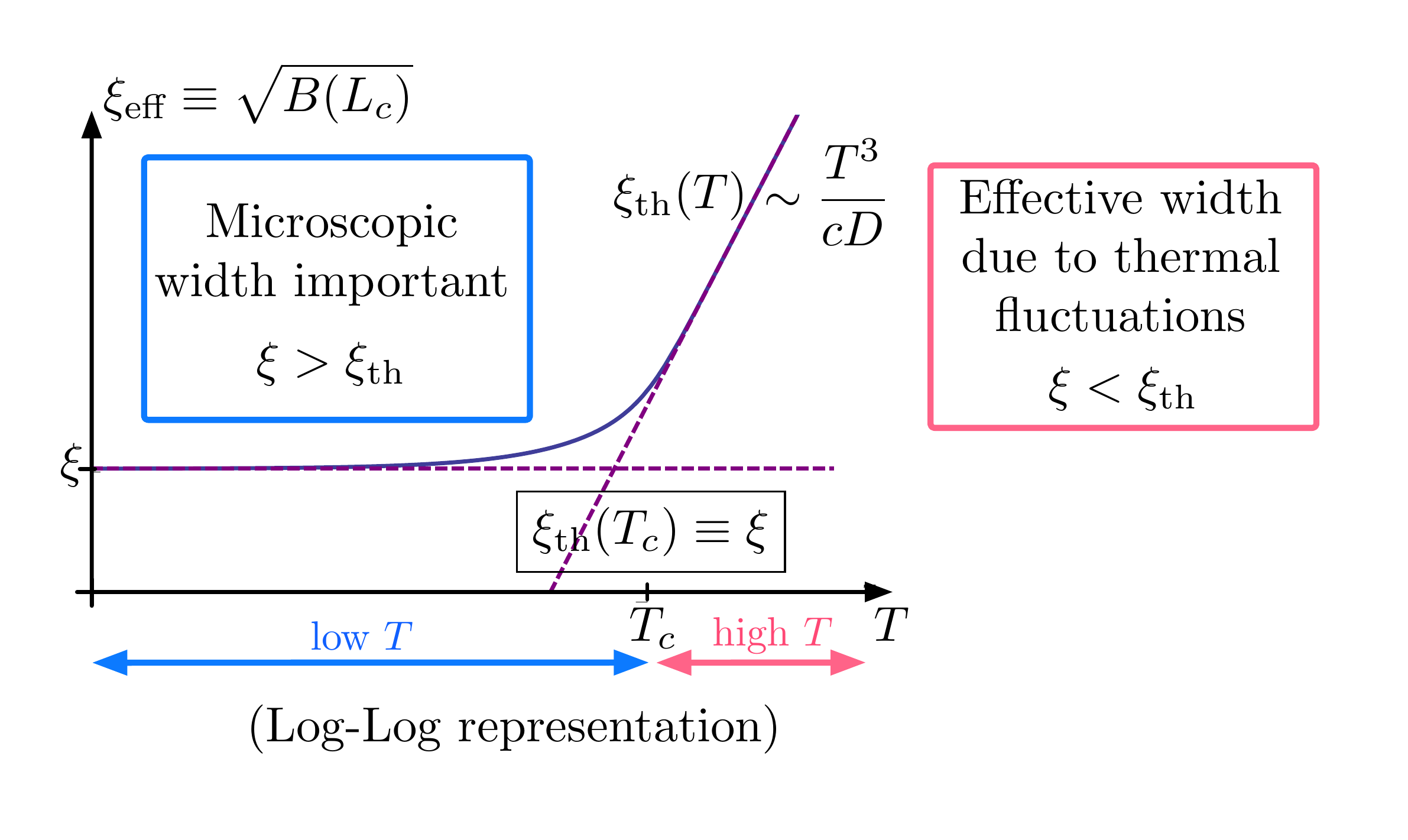}
\caption{Log-Log representation of the effective width $\xi_{\text{eff}}$ of the 1D interface or DP toy model, as a function of temperature. It is defined with respect to the roughness at the beginning of the random-manifold asymptotic regime, i.e. at the Larkin length $L_c$.}
\label{fig-larkin-xieff}
\end{figure}

Plotting this quantity as a function of temperature in a $\log-\log$ representation (cf. Fig.\ref{fig-larkin-xieff}), for both GVM computations, we can clearly observe the existence of the two temperature regimes, by the comparison of the microscopic length $\xi$ and a thermal width ${\xi_{\text{th}} \sim T^3/(cD)}$ as discussed qualitatively with scaling arguments in the section \ref{section-scaling}.
Those two regimes connect when $\xi_{\text{eff}}(T_c) = \xi_{\text{th}}(T_c) = \xi$ defining up to a numerical constant the characteristic temperature ${T_c=(\xi c D)^{1/3}}$.
This result establishes that \textit{the two limits ${\xi \to 0}$ and ${T \to 0}$ cannot be innocently exchanged}, as underlined by the existence of the two opposite FRG regimes of zero-temperature fixed-point \cite{chauve_2000_ThesePC_PhysRevB62_6241} \textit{versus} high-temperature \cite{bustingorry_2010_PhysRevB82_140201}.

The Larkin length $L_c$ is typically below the resolution of the ferromagnetic and ferroelectric DWs experiments mentioned before, and a temperature-dependence study of the (asymptotic) roughness and its \thorn ~exponent has not been carried out on such 1D experimental interfaces yet.
Nevertheless the existence of a low-temperature regime with a physically finite width ${\xi}$ might be relevant even for those experiments, since a crude estimation of $T_c$ for ferromagnetic DWs \cite{repain_2004_EurPhysLett68_460} suggests that it could actually be of the order of magnitude of room temperature \cite{agoritsas_2010_PhysRevB_82_184207}.
As for the phenomenological creep formula discussed previously in the section \ref{section-dynamics},
the whole scaling argument relating the roughness and creep exponents $\zeta$ and $\mu$, which worked so well for ferromagnetic DWs in \cite{lemerle_1998_PhysRevLett80_849}, essentially relies on a comparison of the typical scales at a single characteristic lengthscale $L_c$: a so-called `Larkin domain' encounters a typical energy barrier ${U_c \sim c \xi_{\text{eff}}^2/L_c}$ and it takes a typical minimal force ${F_c \sim c \xi_{\text{eff}}/L_c^2}$ to move it.
Plugging boldly our scaling for $\xi_{\text{eff} (T)}$ and $L_c (T)$ in those expressions, we obtain in the high-$T$ regime that ${U_c \approx T}$ whereas ${F_c \sim T^{-7}}$
and at low-$T$ both those quantities saturate to constants at $T_c$.
However, neither of those predictions are compatible with the affine behavior for ${U_c(T)}$ observed numerically in \cite{kolton_2005_PhysRevLett94_047002}, but at least now we know that this discrepancy might be linked to the appearance at low-$T$ of a whole crossover roughness regime at intermediate lengthscales, thus jeopardising \textit{a priori} any argument based on a single characteristic lengthscale.

\section{Conclusion}
\label{section-conclusion}

We have briefly presented in these notes the physics of disordered elastic systems. 
There are many physical realizations of DES, both interfaces or periodic systems.
In order to probe the disorder-induced glassy properties of a DES, one can study simultaneously its statics via its geometrical fluctuations (roughness and PDF of the relative displacements as a function of the lengthscale)
and its dynamics via its response to an external force (creep, depinning, avalanches, ...).
Different characteristic scales are at stake, and their competition is encoded  at the core of the DES formulation -- the elastic energy $c$, the disorder amplitude $D$, the disorder correlation length $\xi$ and the temperature $T$ --
and mirrored in the lengthscales dependence of observables such as the static roughness or the size distribution of avalanches.

In the low-dimensional case of a 1D interface, an additional physical ingredient \textit{a priori} as innocent as the existence of a finite width $\xi$ turns out to be crucial for the low-temperature properties in statics and \textit{a fortiori} also in the quasistatic dynamics, as emphasized in the FRG `zero-temperature' fixed point of the renormalization approach.
We have shown how a combination of scaling arguments and concrete GVM computations on a pure static observable such as the roughness can already yield non-trivial elements to a coherent picture of the physics at stake.
Application of these ideas in the case of the finite width interface to the dynamical quantities is a very challenging, and experimentally relevant question, that remains to be done.

\section*{Acknowledgements}

This work was supported in part by the Swiss National Science Foundation under MaNEP and Division II.


\tableofcontents

\end{document}